\documentstyle[12pt]{article}
\newcommand{\be}{\begin{equation}}
\newcommand{\ee}{\end{equation}}
\newcommand{\bea}{\begin{eqnarray}}
\newcommand{\eea}{\end{eqnarray}}
\newcommand{\bc}{\begin{center}}
\newcommand{\ec}{\end{center}}
\newcommand{\vac}{\mid 0\rangle}
\newcommand{\x}{{\vec{\rm x}}}
\newcommand{\ka}{{\vec{\rm k}}}
\newcommand{\y}{{\vec{\rm y}}}
\newcommand{\q}{{\vec{\rm q}}}
\newcommand{\p}{{\vec{\rm p}}}

\newcommand{\rv}{{\vec{\rm r}}}

\newcommand{\s}{{\vec{\rm s}}}

\newcommand{\jv}{\vec{J}}
\newcommand{\kap}{\vec{\kappa}}
\newcommand{\di}{\displaystyle \int }
\newcommand{\PP}{{\cal P}}
\newcommand{\Pop}{{\mbox{\bf P}} }

\newcommand{\G}{{\mbox{\bf G}} }
\newcommand{\A}{\tilde{A}}
\newcommand{\B}{\tilde{B}}
\newcommand{\C}{\tilde{C}}
\newcommand{\vc}{{\vec{\rm C}}}
\newcommand%
{\E}%
[1]%
{E_2({\cal P},#1)}%
\newcommand%
{\I}%
[1]%
{{I^{{\cal P}\pm}_{#1}(q)}}%
\newcommand%
{\VF}%
[1]%
{\Phi^{\pm}_{{\cal P}q}(#1)}%
\newcommand%
{\vf}%
[2]%
{\phi^{\pm (#1)}_{{\cal P}q}(#2)_{\alpha\beta}}%
\newcommand%
{\J}%
[1]%
{{\rm J}^{\cal P}_{#1}(\sigma)}%
\newcommand%
{\CC}%
[1]%
{{{\cal C}^{{\cal P}\pm}_{#1}(q)}}%
\newcommand%
{\DD}%
[1]%
{{{\cal D}^{{\cal P}\pm}_{#1}(q)}}%
\newcommand%
{\TT}%
[2]%
{{{\cal T}^{(#1)\pm}_{{\cal P}}(q;#2)}}%
\newcommand%
{\D}%
[1]%
{D^{\cal P}_{#1}(\sigma)}%
\begin{document}
\begin{center}
{\Large \bf Three-particle States in Nonrelativistic Four-fermion Model}
\end{center}

\vspace{4mm}

\begin{center}
A.N.Vall, S.E.Korenblit, V.M.Leviant, D.V.Naumov, A.V.Sinitskaya.
\footnote{This work is partially supported by RFFR N 94-02-05204. } \\
Irkutsk State University, 664003, Gagarin blrd, 20, Irkutsk,
\footnote{E-mail KORENB@math.isu.runnet.ru}
\\ Russia.
\end{center}
\begin{abstract}
{\small
On a nonrelativistic contact four-fermion model
we have shown that the simple $\Lambda$-cut-off prescription together with
definite fine-tuning of the $\Lambda$ dependency of "bare"quantities lead to
self-adjoint semi-bounded Hamiltonian in one-, two- and three-particle
sectors. The fixed self-adjoint extension and exact solutions in two-particle
sector completely define three-particle problem. The renormalized Faddeev
equations for the bound states with Fredholm properties are obtained and
analyzed.}

\end{abstract}
\begin{center}
{\bf 1. Introduction }
\end{center}

Models with contact four-fermion interaction are considered in a
wide range of problems both in solid medium and in quantum field theory.
It is well known that in quantum field theory such interaction is
nonrenormalizable in the frames of conventional perturbation approach.
In our previous works \cite{ALSHT}, \cite{YdFz}, \cite{dnmpZ},
we have demonstrated that nonrelativistic four-fermion quantum field models
possess exact two-particle solutions which clarify the meaning of
renormalization in these models. In the present one we show how these
solutions lead to correct definition of three-particle problem as well.

\begin{center}
{\bf 2. Contact four-fermion models }
\end{center}

Let us consider the following Hamiltonian: $ x=(\x,t=x_0)$,
\bea
&& H =\int d^3x\left\{\Psi^{\dagger a}_{\alpha }(x)\,{\cal E}({\Pop})\,
\Psi^a_{\alpha }(x) -\frac{\lambda}{4}\left[S^2(x)-
{\jv}^2(x)\right]\right\}, \;\;\mbox{ \bf with the conventions:}
\label{H2} \\
&&{\Pop}_{\rm x}=-i\vec{\nabla}_{\rm x};\;\;\;
\epsilon_{\alpha\beta}=-\epsilon_{\beta\alpha};\;\;\;
S(x)=\Psi^{\dagger a}_{\alpha }(x)\,\Psi^a_{\alpha }(x);\;\;\;
{\jv}(x)=(2mc)^{-1}\Psi^{\dagger a}_{\alpha }(x)
\stackrel{\longleftrightarrow}{{\Pop}}\Psi^a_{\alpha }(x);
\nonumber \\
&& \left\{\Psi^a_{\alpha }(x)\,,\Psi^b_{\beta}(y)\right\}\Biggr|_{x_0=y_0}=0;
\;\;\;\;\left\{\Psi^a_{\alpha }(x)\,,\Psi^{\dagger b}_{\beta}(y)\right\}
\Biggr|_{x_0=y_0}=\delta_{\alpha\beta}\delta^{ab}\,\delta_3(\x-\y)
\Longrightarrow\Biggr|_{\x=\y}\delta_{\alpha\beta}\delta^{ab}\;\frac{1}{V^*}.
\nonumber
\eea
Here ${\cal E}(k)$ is arbitrary "bare" one-particle spectrum, $V^*$ has a
meaning of excitation volume, which could be connected with
momentum cut-off $\Lambda $: $6\pi^2/V^*=\Lambda^3$.
This Hamiltonian is invariant under (global) symmetry transformations
$SU_I(2)\times SU_A(2)\times U(1)$.
Introducing Heisenberg fields (HF) in momentum representation
\be
\Psi^a_{\alpha}(\x,t)=\int\frac{d^3k}{(2\pi)^{3/2}}\,e^{i(\ka\x)}\,
b^a_{\alpha}(\ka,t);\;\;
\left\{b^a_{\alpha}(\ka,t)\,,b^{\dagger b}_{\beta}(\q,t)\right\}=
\delta_{\alpha\beta}\delta_{ab}\delta_3(\ka-\q),
\label{HF}
\ee
we consider three different linear operator realizations of HF for $t=0$ via
physical fields, connected by Bogolubov rotations:
$ u_a=\cos \vartheta^a;\;v_a=\sin \vartheta^a;\;(u_a)^2 + (v_a)^2=1,$
\bea
&& b^a_{\alpha}(\ka,0)=e^{\G}\,d^a_{\alpha}(\ka)\,e^{-\G}=
u_a\,d^a_{\alpha}(\ka)-v_a\epsilon_{\alpha\beta}d^{\dagger a}_{\beta}(-\ka),
\nonumber \\
&& \G=\frac{1}{2}\sum_{a=1,2}\vartheta^a\epsilon_{\alpha\beta}\int d^3k
\left[d^{\dagger a}_{\alpha}(\ka)d^{\dagger a}_{\beta}(-\ka)+
d^a_{\alpha}(\ka)d^a_{\beta}(-\ka)\right]=-\G^{\dagger},
\nonumber
\eea
which under condition $u_a v_a=0$ for $a=1,2$ lead to reduced Hamiltonians
in normal form exactly diagonalizable above corresponding vacuums
$d^a_{\alpha}(\ka)\vac=0$:
\bea
&& H=Vw_0+\hat{H};\;\;\;\hat{H}=\hat{H}_0+\hat{H}_I;\;\;\;\hat{H}\{d\}\vac=0;
\;\;\;\left[H\{d\}\,,d^{\dagger a}_{\alpha}(\ka) \right]\vac =
E^a(k)\,d^{\dagger a}_{\alpha}(\ka)\vac ;
\label{dVac} \\
&& w_0=\frac{1}{V^*}\left[\left(2<{\cal E}(k)>-4g\right)
\left(v_1^2+v_2^2\right)-8g (v_1v_2)^2 \right];\quad
<{\cal E}(k)>\stackrel{def}{=}V^*\int \frac{d^3k}{(2\pi)^3}\,{\cal E}(k);
\label{aW0} \\
&& E^a(k)=\frac{g}{(2mc)^2}\left(k^2+<k^2>\right)+g+
(1-2v_a^2)\left[{\cal E}(k)-2g(1+2v_{3-a}^2) \right];
\;\;\;\; <k^2>=\frac{3}{5}\Lambda^2;
\label{Eav} \\
&& \hat{H}_0\{d\}=\sum_{a=1,2}\di d^3k\, E^a(k)\,d^{\dagger a}_{\alpha}(\ka)
\,d^a_{\alpha}(\ka);\;\; \hat{H}_I\{d\}=\sum_{a,b}\di d^3k_1d^3k_2d^3k_3
d^3k_4\;\delta(\ka_1+\ka_2-\ka_3-\ka_4)
\nonumber \\
&&\cdot K^{\PP(ab)}\left(\frac{\ka_1-\ka_2}{2};\frac{\ka_4-\ka_3}{2}\right)
d^{\dagger a}_{\alpha}(\ka_1)d^{\dagger b}_{\beta}(\ka_2)d^b_{\beta}(\ka_3)
d^a_{\alpha}(\ka_4); \qquad g=\frac{\lambda}{4V^*}.
\label{aH0}
\eea
The different realizations correspond to different systems when $v_{1,2}$
independently take values 0,1.
B-system: $v_1=v_2=0$, $d^1_{\alpha}(\ka)=B_{\alpha}(\ka)$,
$d^2_{\alpha}(\ka)=\B_{\alpha}(\ka)$, then $E^{1,2}_B(k)=E_B(k)$.
One can check, that corresponding vacuum state $\vac_B$ is singlet for
both  $SU_I(2)$ and $SU_A(2)$ groups and the one-particle excitations
of $B$ and $\B$ form corresponding fundamental representations.
C-system: $v_1=v_2=1$,
$d^1_{\alpha}(\ka)=\epsilon_{\alpha\beta}C_{\beta}(\ka)$,
$d^2_{\alpha}(\ka)=\epsilon_{\alpha\beta}\C_{\beta}(\ka)$,
$E^{1,2}_C(k)=E_C(k)$.
The symmetry structure of this system is similar to B-system.
A-system: $v_1=0$, $v_2=1$ (or otherwise), which will be considered in more
detail. Let $d^1_{\alpha}(\ka)=A_{\alpha}(\ka)$,
$d^2_{\alpha}(\ka)=\epsilon_{\alpha\beta}\A_{\beta}(\ka)$,
and let $f^{ab}$ be an arbitrary constant $SU_A(2)$ matrix, then for
$ E^{2,1}_A(k)\equiv E^{(+,-)}_A(k)\equiv E_{\A,A}(k)$ corresponding HF
(\ref{HF}) resemble relativistic ones:
\bea
\Psi^a_{\alpha}(x)_A=\int \frac{d^3k}{(2\pi)^{3/2}}
\left[f^{a1}{\cal A}_{\alpha}(\ka,t)e^{-itE^{(-)}_A(k)}+
f^{a2}\tilde{\cal A}^{\dagger}_{\alpha}(-\ka,t)e^{itE^{(+)}_A(-k)}\right]
e^{i(\ka\x)},
\nonumber \\
\mbox{where: }\left\{ \begin{array}{c}
{\cal A}_{\alpha}(\ka,t)\, e^{-itE^{(-)}_A(k)} \\
\tilde{\cal A}_{\alpha}(\ka,t)\, e^{-itE^{(+)}_A(k)}
\end{array} \right\} =e^{iHt}
\left\{\begin{array}{c} A_{\alpha}(\ka) \\ \A_{\alpha}(\ka) \end{array}
\right\} e^{-iHt}.\qquad \qquad\quad\qquad\qquad
\label{GFA}
\eea
It is a simple matter to show that for system A the $SU_A(2)$ and $U(1)$
symmetries turn out to be spontaneously broken and there are four composite
Goldstone states \cite{ALSHT}, \cite{YdFz}.

\begin{center}
{\bf 3. Two-particle eigenvalue problems}
\end{center}

The interaction between all particles in the systems B and C is the same
as for $AA$, $\A\A$ in system A. So it is enough to consider the later
one. Hereafter $BB$ means $BB$, $\B\B$, $B\B$ and analogously for $CC$.
Defining the two-particle interaction kernels and energies as:
\bea
&& K^{\PP(QQ')}(\s,\ka)=\left\{\begin{array}{cc}
K^{\PP\{+\}}(\s,\ka), & \mbox{for} \quad QQ'= \A\A,AA,BB,CC \\
K^{\PP\{-\}}(\s,\ka), & \mbox{for} \quad QQ'= A\A \end{array}
\right.,\;\mbox{ where,}
\nonumber \\
&& -2K^{\PP\{\pm\}}(\s,\ka)=\frac{V^*}{(2\pi)^3}\cdot\frac{2g}{(2mc)^2}
\left[(\s+\ka)^2-\PP^2 \pm (2mc)^2\right];\;\mbox{ as well as:}
\label{K2} \\
&& E^{QQ'}_2(\PP,\ka)\equiv E_{Q}\left(\frac{\PP}{2}+\ka\right)+
E_{Q'}\left(\frac{\PP}{2}-\ka\right)= E^{\{\pm \}}_2(\PP,\ka);\quad
\mbox{ so: }
\label{E2+} \\
&& E^{\{+\}}_2(\PP,\ka)=\frac{2g}{(2mc)^2}\left[<k^2> + (2mc)^2+k^2+
\frac{\PP^2}{4}\right]+
\nonumber \\
&& +\left\{\!\begin{array}{c}\displaystyle \!\!\!\!
\pm \left[4g-{\cal E}(\frac{\PP}{2}+\ka)-{\cal E}(\frac{\PP}{2}-\ka)\right]\\
\\ \displaystyle \,
\;\pm\left[12g-{\cal E}(\frac{\PP}{2}+\ka)-{\cal E}(\frac{\PP}{2}-\ka)\right]
\end{array}\! \right\}, \;\;\;\mbox{ for }
QQ'= \left\{\!\!\begin{array}{c}
\left[\!\!\begin{array}{c} \A\A \\  BB \!\!\end{array}\right] \\
\left[\!\!\begin{array}{c} CC \\  AA \!\!\end{array}\right]\\
\end{array}\!\!\right\};
\nonumber \\
&& E^{\{-\}}_2(\PP,\ka)=\frac{2g}{(2mc)^2}\left[<k^2>- (2mc)^2+k^2+
\frac{\PP^2}{4}\right]+\left[{\cal E}(\frac{\PP}{2}+\ka)-
{\cal E}(\frac{\PP}{2}-\ka)\right],
\mbox{ for } QQ'=A\A,
\nonumber
\eea
we can formulate both scattering and bound state two-particle eigenvalue
problems in the Fock eigenspace of kinetic part $\hat{H}_0$ of the reduced
Hamiltonian $\hat{H}$ (\ref{dVac}), (\ref{aH0}):
\bea
&& \hat{H}\mid R^{\pm (QQ')}_{\alpha\beta}(\PP,\q)\rangle=E^{QQ'}_2(\PP,\q)\,
\mid R^{\pm (QQ')}_{\alpha\beta}(\PP,\q)\rangle ;\qquad
\hat{H}\mid {\rm B}^{\PP(QQ')}_{\alpha \beta} \rangle  =
M^{QQ'}_2(\PP)\mid {\rm B}^{\PP(QQ')}_{\alpha \beta} \rangle ;
\label{egnvl} \\
&& \mid R^{\pm (QQ')}_{\alpha\beta}(\PP,\q)\rangle =\di d^3k\,
\Phi^{\pm QQ'}_{\PP q}(\ka)\mid R^{0(QQ')}_{\alpha\beta}(\PP,\ka)\rangle;
\;\;\mid {\rm B}^{\PP(QQ')}_{\alpha \beta}\rangle = \int d^3k\,
\Phi^{QQ'}_{\PP b}(\ka)\mid R^{0(QQ')}_{\alpha\beta}(\PP,\ka)\rangle;
\nonumber \\
&& \mid R^{0(QQ')}_{\alpha\beta}(\PP,\ka)\rangle =
\hat{Q}^{\dagger}_{\alpha }(\frac{\PP}{2}+\ka)\,
\hat{Q}'^{\dagger}_{\beta }(\frac{\PP}{2}-\ka) \vac ;\qquad
M^{QQ'}_2(\PP)=E^{QQ'}_2(\PP,q=ib);
\label{egnst}
\eea
($\hat{Q},\hat{Q}'$ stands for creation operators $A,\A$, or $B,\B$, or
$C,\C$) in terms of Schroedinger equation for corresponding
(scattering or bound state) wave function:
\be
\left[E^{QQ'}_2(\PP,\ka)-M^{QQ'}_2(\PP)\right]\Phi^{QQ'}_{\PP b}(\ka)=
-2\int d^3s\,\Phi^{QQ'}_{\PP b}(\s)\, K^{\PP(QQ')}(\s,\ka).
\label{Prblm}
\ee
As was shown in \cite{ALSHT}, \cite{YdFz}, this equation realy has
for $\{-\}$ case exact simple solutions corresponding to the Goldstone states
almost independently from the very form of "bare" spectrum.

For the $AA$ or $\A\A$ two-particle states (case $\{+\}$) the quadratic
form of "bare" spectrum according to (\ref{dVac}), (\ref{Eav}) transforms to
the renormalized one:
\bea
&& {\cal E}(k)=\frac{k^2}{2m}+{\cal E}_0\,\longrightarrow\,
E^{(\pm)}_A(k)=\frac{k^2}{2{\cal M}^{(\pm)}}+E^{(\pm)}_{A0};\quad\;
\frac{1}{2{\cal M}^{(\pm)}}=\frac{g}{(2mc)^2}\mp \frac{1}{2m};
\label{bare} \\
&& E^{(\pm)}_{A0}=g\left(\frac{<k^2>}{(2mc)^2}-1\pm 4\right)\mp {\cal E}_0;
\quad\;\lambda_0=\frac{\lambda {\cal M}^{(\pm)}}{2};\quad\;
\mu_0=\frac{\lambda_0}{(2mc)^2}.
\label{mug}
\eea
The eq.(\ref{Prblm}) in configuration space reveals strongly singular
interaction potential, studied in \cite{BrFdd}, \cite{YShr}, \cite{Shond}:
\bea
\left(-\nabla^2_{\rm x}-q^2\right)\psi_q(\x)=\delta_3(\x)R_1(q)
-\nabla^2_{\rm x}\delta_3(\x)R_2(q)-2\mu_0
\left((\vec{\nabla}\psi_q)(0)\cdot\vec{\nabla}_{\rm x}\delta_3(\x)\right);
\label{Scrd2} \\
R_1(q)\equiv (\lambda_0-\mu_0\PP^2)\psi_q(0)-\mu_0(\nabla^2\psi_q)(0);\;\;\;
R_2(q)\equiv\mu_0\psi_q(0).
\label{R12}
\eea
The first and second terms in the R.H.S. of (\ref{Scrd2}) represent
interaction with the orbital momentum $l=0$, the third one gives interaction
only for $l=1$ and disappear after integration over unit sphere of
$\q$-directions. Among the obtained in \cite{ALSHT},\cite{YdFz} different
solutions for the two-particle wave functions (\ref{egnvl}) which correspond
to different self-adjoint extensions \cite{Shond} of operator (\ref{Scrd2}),
(\ref{R12}), the use of $\Lambda$-cut-off regularization \cite{BrFdd}
together with simple subtraction procedure, as well as in \cite{Ber},
with $\Lambda\rightarrow\infty$ pick out the following renormalized ones:
\bea
\mid l,{\rm J,m};\PP,q\rangle^{\pm}=\di d^3k\,\vf{l,\rm J,m}{\ka}\,
\mid R^{0(QQ')}_{\alpha\beta}(\PP,\ka )\rangle;\qquad
\vf{l,\rm J,m}{\ka}=\chi^{(\rm J,m)}_{\alpha \beta}\,
\Phi^{\pm(l)}_{{\cal P}q}(\ka);\quad
\label{2states} \\
\mbox{where for } Q=Q':\quad
\vf{l,\rm J,m}{\ka}=-\phi^{\pm (l,{\rm J,m})}_{{\cal P}q}
(-\ka)_{\beta\alpha};\; \mbox{ and then }\;l={\rm J}=0,1;
\qquad\qquad
\nonumber \\
\chi^{(0,0)}_{\alpha \beta}=\frac{1}{\sqrt{2}}
(\delta_{\alpha 1}\delta_{\beta 2}-\delta_{\alpha 2}\delta_{\beta 1});\;\;\;
\chi^{(1,0)}_{\alpha \beta}=\frac{1}{\sqrt{2}}
(\delta_{\alpha 1}\delta_{\beta 2}+\delta_{\alpha 2}\delta_{\beta 1});\;\;\;
\chi^{(1,\pm 1)}_{\alpha \beta}=\left\{
\begin{array}{c}\delta_{\alpha 1}\delta_{\beta 1} \\
\delta_{\alpha 2}\delta_{\beta 2} \end{array}\right.;\qquad
\label{SpSt} \\
\Phi^{\pm(l)}_{{\cal P}q}(\ka)=\frac{1}{2}\left[\delta_3(\ka-\q)+
(-1)^{l} \delta_3(\ka+\q)\right]+\frac{\TT{l}{k} }{k^2-q^2\mp i0 };
\qquad \Phi^{(0)}_{\PP b}(\ka)=\frac{const}{k^2+b^2};\qquad
\label{Phil} \\
\TT{0}{k}=\left.\mu_0 \frac{(2mc)^2+\gamma <k^2>-\PP^2+q^2+(1-\gamma)k^2}
{(2\pi)^3\left[{\cal D}^{\PP}(\mp iq)-{\cal D}^{\PP}(b)\right]}
\right|_{\Lambda\rightarrow\infty}\Longrightarrow
\frac{-\left(2\pi^2\right)^{-1}\Upsilon}{(\Upsilon+b\mp iq)(b\pm iq)};\ \
\label{T0rnrm} \\
{\cal D}^{\PP}(\varrho)\equiv(\gamma-1)^2-{\rm I}_0(\varrho)\left[
(2mc)^2+\gamma <k^2>-\PP^2-(2-\gamma)\varrho^2\right];
\qquad {\cal D}^{\PP}(b)=0;\qquad
\label{D02} \\
\TT{1}{k}=\left.\frac{2\mu_0(\ka\cdot\q)}{(2\pi)^3}\cdot\left[1-
\frac{2}{3}{\rm I}_1(\mp iq)\right]^{-1}\right|_{\Lambda\rightarrow\infty}
\Longrightarrow 0; \quad {\rm I}_n(\varrho)=
\mu_0\di\frac{d^3k}{(2\pi)^3}\cdot\frac{\left(k^2\right)^n}{k^2+\varrho^2};
\quad
\label{T1}\\
\mbox{where }\; g=\Lambda^2G(\Lambda),\;\;(2mc)^2=\Lambda^2\nu(\Lambda),
\;\;{\cal E}_0=\Lambda^2\epsilon(\Lambda),\;\;
\gamma^{(\pm)}(\Lambda)=\frac{\mu_0}{V^*}=1\pm\frac{{\cal M}^{(\pm)}}{m},
\mbox{ with} \qquad
\label{gamma} \\
G(\Lambda)=G_0+G_1/\Lambda+G_2/\Lambda^2+\ldots;\mbox{ and so for }
\nu(\Lambda),\,\epsilon(\Lambda),\,\gamma(\Lambda); \mbox{ and if }\;
G_0,\nu_0\neq 0, \mbox{ then }
\nonumber \\
\mbox{one has: }
\gamma^{(\pm)}_0=\gamma_0=1;\;\;\gamma^{(\pm)}_1=
\pm c\frac{\sqrt{\nu_0}}{G_0};\;\;
{\cal M}^{(\pm)}_0={\cal M}_0=\frac{\nu_0}{2G_0};\;\;
\Upsilon=\frac{\pi}{2}\left(\frac{3}{5}\gamma_1+\nu_1\right);\;\;
\nu_0=-\frac{3}{5}.\
\label{01}
\eea
The last equality in (\ref{01}) reflects the bound state condition
(\ref{D02}) which serves here as a dimensional transmutation condition
\cite{YdFz}, \cite{Thorn}. Thus $\Upsilon$ and real $b$ are
{\it arbitrary} parameters of self-adjoint extension \cite{Shond}
which may be formally partially expressed via parameters of $\Lambda$
-dependence of "bare" quantities by fine tuning relations (\ref{01}).
Strictly speaking, the solution (\ref{Phil}), (\ref{T0rnrm}), (\ref{D02})
imply a self-adjoint extension for restricted on appropriate subspace of
$L^2$ initial free Hamiltonianto to extended Hilbert space $L^2\oplus C^1$.
The additional discrete component of eigenfunctions "improves" their
scalar product, it is completely defined by the same parameters of self-
adjoint extension but does not affect on physical meaning of obtained
solution in ordinary space \cite{Shond}, \cite{Fewst}, \cite{Shond3}.

Another extension corresponds to the choice of finite "bare" mass that is
possible only for B-system and for $(-)$ case of A-system. Thus
$G_{0,1}=\nu_{0,1}=0$, and (\ref{gamma}) with transmutation condition
(\ref{D02}) leads to the solution which coincides with the well known
extension in $L^2$ \cite{BrFdd} for operator (\ref{Scrd2}) with
$\mu_0\equiv 0$, for which: $\gamma^{(-)}_0=1-(3/4)(3\pm\sqrt{5}) <1$,
and $(r=|\x|)$
\be
{\cal M}^{(-)}_0=m\frac{3}{4}(3\pm\sqrt{5});\;\;\;
\TT{0}{k}\mid_{\Lambda\rightarrow\infty}=
\frac{-\left(2\pi^2\right)^{-1}}{(b\pm iq)};\;\;\;
\psi^{(0)}_b(\x)=\frac{\sqrt{8\pi b}}{4\pi}\frac{e^{-br}}{r}.
\label{FB}
\ee

\begin{center}
{\bf 4. Three-particle eigenvalue problems}
\end{center}

From Schroedinger equation with Hamiltonian (\ref{aH0}) for eigenstate of
three identical (A) particles with total momentum $\PP$ follows equation
for wave function satisfying Pauly principle: (further $E_A(\ka)=E(\ka)$)
\bea
|3,\PP\rangle =
\int d^3q_1 d^3q_2 d^3q_3 D^{(\PP,{\rm J,m})}_{\alpha\beta\gamma}
(\q_1\q_2\q_3)A^{\dagger}_{\alpha}(\q_1) A^{\dagger}_{\beta}(\q_2)
A^{\dagger}_{\gamma}(\q_3)\vac ;\;\;\;
\hat{H}|3,\PP\rangle =M_3(\PP)|3,\PP\rangle ;
\label{3st} \\
\left. D^{(\PP,{\rm J,m})}_{\alpha\beta\gamma}(\q_1\q_2\q_3)=-
D^{(\PP,{\rm J,m})}_{\beta\alpha\gamma}(\q_2\q_1\q_3)\equiv
\delta(\q_1+\q_2+\q_3-\PP)\,\tilde{\cal D}^{(\PP,{\rm J,m})}
_{\alpha\beta\gamma}(\q_1\q_2\q_3)\right|_{\q_1+\q_2+\q_3=\PP};
\label{prop} \\
D^{(\PP,{\rm J,m})}_{\alpha\beta\gamma}(\q_1\q_2\q_3)
\left[\sum_{i=1}^3 E(\q_i)-M_3(\PP)\right]=-\int d^3k_1 d^3k_2 d^3k_3
D^{(\PP,{\rm J,m})}_{\alpha\beta\gamma}(\ka_1\ka_2\ka_3)
{\cal H}(\ka_1 \ka_2 \ka_3|\q_1\q_2\q_3);
\label{3Deq} \\
{\cal H}(\ka_1\ka_2\ka_3|\q_1\q_2\q_3) =\frac{-\lambda}{2(2\pi)^3}\;
\delta\Bigl(\sum^3_{i=1}\ka_i-\sum^3_{i=1}\q_i\Bigr)
\left\{\sum^3_{1=n\neq j<l}\delta(\ka_n-\q_n)\left[1-
\frac{(\ka_j+\q_j)\cdot(\ka_l+\q_l)}{(2mc)^2}\right]\right\}.
\label{3kern}
\eea
The kernel (\ref{3kern}) obviously reproduces all permutation symmetries and
momentum conservation. So, it seems convenient to simplify the study of
spin-symmetry structure of wave function using the formal function of three
(dependent) variables like
$\tilde{\cal D}^{(\PP,{\rm J,m})}_{\alpha\beta\gamma}(\q_1\q_2\q_3)$
(\ref{prop}) and introducing corresponding "formfactors":
\be
\left.\tilde{\cal K}^{(\PP,{\rm J,m})}_{\alpha\beta\gamma}(\q_1\q_2\q_3)
\right|_{\q_1+\q_2+\q_3=\PP}=\left[\sum_{i=1}^3 E(\q_i)-M_3(\PP)\right]\left.
\tilde{\cal D}^{(\PP,{\rm J,m})}_{\alpha\beta\gamma}(\q_1\q_2\q_3)
\right|_{\q_1+\q_2+\q_3=\PP}.
\label{3KKK}
\ee
Since the momentum conservation condition is totally symmetrical over $\q_j$,
the $\tilde{\cal K}$ and $\tilde{\cal D}$ have the same spin-symmetry
structure as of $D$. Namely, let $\left([\ldots]\right) \{\ldots\}$ means
hereafter (anti) symmetrization over internal indices, then one has:
\bea
&& \tilde{\cal K}^{(\PP,1/2,{\rm m})}_{(X)\alpha\beta\gamma}
(\q_1\q_2\q_3)=
\Gamma^{1/2,{\rm m}}_{\alpha\{\beta\gamma\}}X(\q_1[\q_2\q_3])+
\Gamma^{1/2,{\rm m}}_{\gamma\{\alpha\beta\}}X(\q_3[\q_1\q_2])+
\Gamma^{1/2,{\rm m}}_{\beta\{\gamma\alpha\}}X(\q_2[\q_3\q_1])=
\label{K1/2X} \\
&& =\Gamma^{1/2,{\rm m}}_{\alpha\{\beta\gamma\}}
K^{(\PP)}_X(\{\q_1\q_2\}\q_3)-\Gamma^{1/2,{\rm m}}
_{\gamma\{\alpha\beta\}}K^{(\PP)}_X(\{\q_2\q_3\}\q_1);\;\;
K^{(\PP)}_X(\{\q_1\q_2\}\q_3)\equiv X(\q_1[\q_2\q_3])+X(\q_2[\q_1\q_3]);
\nonumber \\
&& \tilde{\cal K}^{(\PP,1/2,{\rm m})}_{(Y)\alpha\beta\gamma}
(\q_1\q_2\q_3)=
\Gamma^{1/2,{\rm m}}_{\alpha [\beta\gamma]}Y(\q_1\{\q_2\q_3\})+
\Gamma^{1/2,{\rm m}}_{\gamma [\alpha\beta]}Y(\q_3\{\q_1\q_2\})+
\Gamma^{1/2,{\rm m}}_{\beta [\gamma\alpha]}Y(\q_2\{\q_3\q_1\})=
\label{K1/2Y} \\
&& =\Gamma^{1/2,{\rm m}}_{\alpha [\beta\gamma]}
K^{(\PP)}_Y([\q_1\q_2]\q_3)+\Gamma^{1/2,{\rm m}}_
{\gamma [\alpha\beta]}K^{(\PP)}_Y([\q_3\q_2]\q_1);\quad
K^{(\PP)}_Y([\q_1\q_2]\q_3)\equiv Y(\q_1\{\q_2\q_3\})-Y(\q_2\{\q_1\q_3\});
\nonumber \\
&& \tilde{\cal K}^{(\PP,3/2,{\rm m})}_{\alpha\beta\gamma}(\q_1\q_2\q_3)=
\Gamma^{3/2,{\rm m}}_{\{\alpha\beta\gamma\}}K^{(\PP)}_Z([\q_1\q_2\q_3]).
\label{K3/2}
\eea
Here the following properties of three-spin-wave functions were used:
\bea
&& \Gamma_{\alpha\beta\gamma}^{1/2,1/2}=
a\delta_{\alpha 2}\delta_{\beta 1} \delta_{\gamma 1} +
b\delta_{\alpha 1}\delta_{\beta 2} \delta_{\gamma 1} +
c\delta_{\alpha 1}\delta_{\beta 1} \delta_{\gamma 2},\mbox{ with }a+b+c=0,
\mbox{ what means, that}
\label{gam0} \\
&& \Gamma_{\alpha\beta\gamma}^{1/2,{\rm m}}+
\Gamma_{\gamma\alpha\beta}^{1/2,{\rm m}}+
\Gamma_{\beta\gamma\alpha}^{1/2,{\rm m}}=0;\;\;\;
\Gamma_{\{\alpha\beta \gamma\}}^{3/2,3/2}=
\delta_{\alpha 1} \delta_{\beta 1} \delta_{\gamma 1};\;\;
\Gamma_{\{\alpha\beta\gamma\}}^{3/2,1/2}=
\delta_{\alpha 2}\delta_{\beta 1}\delta_{\gamma 1}+
\delta_{\alpha 1}\delta_{\beta 2}\delta_{\gamma 1}+
\delta_{\alpha 1}\delta_{\beta 1}\delta_{\gamma 2}.
\nonumber
\eea
To change projection m on $-$m it is enough to replace indices
$1\leftrightarrow 2$. For case J=1/2 three-spin-functions with definite
partial symmetry correspond to eigenvalue of definite spin-permutation
operator: $\Sigma_{23}=+1,\;(X),\;b=c,\;a=-2c $, for symmetrical function
$\Gamma^{1/2,{\rm m}}_{\alpha\{\beta\gamma\}}$; and
$\Sigma_{23}=-1,\;(Y),\;b=-c,\;a=0 $, for antisymmetrical one
$\Gamma^{1/2,{\rm m}}_{\alpha [\beta\gamma]}$.
All the "formfactors" satisfy the same equation and differ only by the
symmetry type $S=X,Y,Z$:
\be
K^{(\PP)}_S(\q_1\q_2\q_3)=-\di d^3k_1 d^3k_2 d^3k_3
\frac{K^{(\PP)}_S(\ka_1\ka_2\ka_3)}
{\sum_{i=1}^3 E(\ka_i)-M_3(\PP)}{\cal H}(\ka_1\ka_2\ka_3|\q_1\q_2\q_3).
\label{KSgen}
\ee
Now we put consequently for every term of the kernel (\ref{3kern}):
\bea
&& 1,2,3=n\neq j\neq l,\;\;j<l;\quad\;\ka_j-\ka_l=2\s,\;\;\;
\ka_j+\ka_l=\rv_n,\;\mbox{ with }\;\; \q_1+\q_2+\q_3=\PP,
\nonumber \\
&& \mbox{obtaining:  }\;\rv_n=\PP-\q_n;\quad \;(\ka_j+\q_j)\cdot
(\ka_l+\q_l)=(\PP-\q_n)^2-\left(\s+\frac{\q_j-\q_l}{2}\right)^2,
\nonumber
\eea
and thus immediately find out the general structure of "formfactors"
in (\ref{KSgen}):
\bea
K^{(\PP)}_S(\q_1\q_2\q_3)=\sum^3_{1=n\neq j<l}\left[
 \vc_{Sn}(\q_n)\cdot (\q_j-\q_l)+A_{Sn}(\q_n)+B_{Sn}(\q_n)(\q_j-\q_l)^2
\right],\quad\mbox{ with}
\label{Kstrct} \\
\left.\begin{array}{c} A_{Sn}(\q) \\ B_{Sn}(\q) \\
\vc_{Sn}(\q) \end{array}\right\}=
\frac{\lambda}{2(2mc)^2} \di \frac{d^3s}{(2\pi)^3}\cdot\frac{
K^{(\PP)}_S(\ka_1\ka_2\ka_3)}{E(\ka_n)+E(\ka_j)+E(\ka_l)-M_3(\PP)}\cdot
\left\{\begin{array}{c}
(2mc)^2+\s^2-(\PP-\q)^2 \\ 1/4  \\ \s \end{array}\!\!\! \right.,
\nonumber \\
\mbox{where for }\; 1,2,3=n\neq j\neq l,\;\;j<l:\;\; \ka_n=\q;\;\;
\ka_j=\frac{\PP-\q}{2}+\s\equiv\kap_+;\;\;
\ka_l=\frac{\PP-\q}{2}-\s\equiv\kap_-.
\label{ABC}
\eea
The system of coupled integral equations (\ref{Kstrct}), (\ref{ABC}) may
be essentially simplified by utilizing the symmetry of functions
$K^{(\PP)}_S$ (\ref{K1/2X}), (\ref{K1/2Y}), (\ref{K3/2}), that demands
for example:
\bea
&& \vc_{Z1}(\q)=-\vc_{Z2}(\q)=\vc_{Z3}(\q)\equiv\vc_{Z}(\q);\quad
A_{Zj}(\q)=B_{Zj}(\q)\equiv 0; \;\mbox{ and thus:}
\nonumber \\
&& K^{(\PP)}_Z([\q_1\q_2\q_3])=\vc_Z(\q_1)\cdot (\q_2-\q_3)+
\vc_Z(\q_2)\cdot (\q_3-\q_1)+\vc_Z(\q_3)\cdot (\q_1-\q_2).
\label{KZ} \\
&& \mbox{Analogously: }\;\;A_{X1}(\q)=A_{X2}(\q)\equiv A_{X}(\q);\quad
B_{X1}(\q)=B_{X2}(\q)\equiv B_{X}(\q);
\nonumber \\
&& \vc_{X1}(\q)=\vc_{X2}(\q)\equiv\vc_{X}(\q);\quad
\vc_{X3}(\q)=0;\quad Q_{Xj}(\q;\p)\equiv A_{Xj}(\q)+\p^2 B_{Xj}(\q);
\nonumber \\
&& K^{(\PP)}_X(\{\q_1\q_2\}\q_3)=Q_{X}(\q_1;\q_2-\q_3)
+Q_{X}(\q_2;\q_1-\q_3)+Q_{X3}(\q_3;\q_1-\q_2)+
\nonumber \\
&& +\vc_X(\q_1)\cdot (\q_2-\q_3)+\vc_X(\q_2)\cdot (\q_1-\q_3).
\label{KX} \\
&& A_{Y1}(\q)=-A_{Y2}(\q)\equiv A_{Y}(\q);\quad
B_{Y1}(\q)=-B_{Y2}(\q)\equiv B_{Y}(\q);\quad A_{Y3}(\q)=B_{Y3}(\q)=0;
\nonumber \\
&& \vc_{Y1}(\q)=-\vc_{Y2}(\q)\equiv\vc_{Y}(\q);\quad
\vc_{Y3}(\q)\equiv 2\left(\vc_{Y0}(\q)-\vc_{Y}(\q)\right);
\nonumber \\
&& K^{(\PP)}_Y([\q_1\q_2]\q_3)=Q_{Y}(\q_1;\q_2-\q_3)-
Q_{Y}(\q_2;\q_1-\q_3)+\vc_{Y3}(\q_3)\cdot (\q_1-\q_2)+
\nonumber \\
&& +\vc_Y(\q_1)\cdot (\q_2-\q_3)-\vc_Y(\q_2)\cdot (\q_1-\q_3).
\label{KY}
\eea
Solving now every of these systems (\ref{ABC}) with (\ref{KZ}), (\ref{KX}),
(\ref{KY}), as nonhomogeneous algebraic one, where unknown integral terms
would be considered as free ones, we come to corresponding homogeneous
systems of Faddeev integral equations:
\bea
&& \vc_Z(\q)=\di\frac{d^3s}{(2\pi)^3}\cdot \frac{1}{(s^2+\varrho^2)}
\left[\frac{-2\mu_0\;\s}{1-\frac{2}{3}{\rm I_1(\varrho )}} \right]
\vc_Z(\kap_+)\cdot (\q-\kap_-);\;\mbox{ and the same eq. for }\;\vc_{Y0}(\q).
\qquad
\label{CZ} \\
&& \mbox{Let: }
Q_{X0}(\q;\p)=Q_{X3}(\q;\p)+2Q_{X}(\q;\p);\qquad \Delta_{X}(\q;\p)
=Q_{X}(\q;\p)-Q_{X3}(\q;\p);\quad\mbox{then:}
\nonumber \\
&& Q_{X0}(\q;2\rv)=\di\frac{d^3s}{(2\pi)^3}\cdot
\frac{2\mu_0}{(s^2+\varrho^2)}\left[\frac{{\cal O}^{\PP}(\q;\s,\rv)}
{{\cal D}^{\PP-q}(\varrho )}\right]Q_{X0}(\kap_+;\q-\kap_-);
\qquad
\label{QX} \\
&& \Delta_{X}(\q;2\rv)=\di\frac{d^3s}{(2\pi)^3}\cdot
\frac{(-\mu_0)}{(s^2+\varrho^2)}\left[\frac{{\cal O}^{\PP}(\q;\s,\rv)}
{{\cal D}^{\PP-q}(\varrho )}\right]\left(\Delta_{X}(\kap_+;\q-\kap_-)
-\vc_X(\kap_+)\cdot (\q-\kap_-)\right);
\nonumber \\
&& \vc_X(\q)=\di\frac{d^3s}{(2\pi)^3}\cdot \frac{1}{(s^2+\varrho^2)}
\left[\frac{\mu_0\;\s}{1-\frac{2}{3}{\rm I_1(\varrho)}} \right]
\left(\vc_X(\kap_+)\cdot (\q-\kap_-)+\Delta_{X}(\kap_+;\q-\kap_-)\right).
\nonumber \\
&& \vc_Y(\q)-\vc_{Y0}(\q)=\di\frac{d^3s}{(2\pi)^3}\cdot\frac{1}{(s^2+\varrho^2)}
\left[\frac{\mu_0\;\s}{1-\frac{2}{3}{\rm I_1(\varrho)}} \right]
\left(\vc_Y(\kap_+)\cdot (\q-\kap_-)-Q_{Y}(\kap_+;\q-\kap_-)\right);
\nonumber \\
&& Q_{Y}(\q;2\rv)=\di\frac{d^3s}{(2\pi)^3}\cdot
\frac{(-\mu_0)}{(s^2+\varrho^2)}\left[\frac{{\cal O}^{\PP}(\q;\s,\rv)}
{{\cal D}^{\PP-q}(\varrho) }\right]\left(Q_{Y}(\kap_+;\q-\kap_-)+
3\vc_Y(\kap_+)\cdot (\q-\kap_-)-\right.
\label{QY} \\
&& \left.-2\vc_{Y0}(\kap_+)\cdot (\q-\kap_-)\right).\quad \mbox{ Here: }\;\;
{\cal O}^{\PP}(\q;\s,\rv)\equiv (2mc)^2+\gamma <k^2>-(\PP-\q)^2-\varrho^2+
\nonumber \\
&& +(1-\gamma)(s^2+r^2+\varrho^2)+{\rm I}_0(\varrho)(s^2+\varrho^2)
(r^2+\varrho^2);\quad\;
\omega^2(\PP)={\cal M}_0\Bigl(3E_0-M_3(\PP)\Bigr);
\nonumber \\
&& E(\q)+E(\kap_+)+E(\kap_-)-M_3(\PP)\equiv \frac{s^2+\varrho^2}{{\cal M}_0};
\;\;\;\varrho^2=\varrho^2(q)\equiv \frac{3}{4}q^2+\frac{\PP^2}{4}-
\frac{(\q\PP)}{2}+\omega^2(\PP).
\nonumber
\eea
For finite $\Lambda$ one can easily recognize the interior of square brackets
in the kernels of that equations as off-shell extensions of (half-off-shell)
two-particle T-matrices from L.H.S. of (\ref{T0rnrm}), (\ref{T1}). However,
for $\Lambda\rightarrow\infty$ the all these off-shell T-matrices obviously
coincide with the corresponding on-shell renormalized ones given by R.H.S. of
(\ref{T0rnrm}), (\ref{T1}).
So, when $\Lambda\rightarrow\infty$, one observes here, as well as in
two-particle case \cite{YdFz}, the restoration of Galileo invariance,
and comes to further simplifications $\vc_{X,Y,Z}=B_{X,Y}=0$ resulting to
renormalized equations for the functions of one variable:
\bea
&& Q_{X0}(\q)=\frac{T(\varrho(q))}{\pi^2}\di d^3s
\frac{Q_{X0}(\kap_+)}{(s^2+\varrho^2(q))};
\label{QX0} \\
&& \left.\begin{array}{c} Q_Y(\q) \\ \Delta_X(\q) \end{array}\right\}=
-\frac{T(\varrho(q))}{2\pi^2}\di d^3s
\frac{1}{(s^2+\varrho^2(q))}\left\{
\begin{array}{c} Q_Y(\kap_+) \\ \Delta_X(\kap_+)\end{array}\right. ;
\label{QDY} \\
&& T(\varrho)\equiv 2\pi^2 \TT{0}{q}\Bigr|_{q=\pm i\varrho}
 =\frac{\Upsilon}{(\Upsilon+b+\varrho)(\varrho-b)}\,
\sim\,\frac{\Upsilon}{\varrho^2}\Bigr|_{\varrho \rightarrow \infty}.
\label{ATgn}
\eea
Obviously, the eqs. (\ref{QX0}) and (\ref{QDY}) can not have nontrivial
solutions simultaneously. Therefore, two different possibilities appear: \\
1) eq. (\ref{QX0}) has nonzero solution. Then $Q_Y(\q)=\Delta_X(\q)=0$,
$Q_{X}(\q)=A_X(\q)$; $K^{(\PP)}_Z=K^{(\PP)}_Y=0$,
$K^{(\PP)}_X(\{\q_1\q_2\}\q_3)= A_X(\q_1)+A_X(\q_2)+A_X(\q_3)$,
where $A_X(\q)$ satisfies the eq. (\ref{QX0}) which coincides with
Shondin's equation \cite{Shond3}. \\
2) eq. (\ref{QDY}) has nonzero solution. Then $Q_{X0}=0$, and coinciding
equations on $\Delta_X(\q)$ and $Q_Y(\q)$  define in principle
the coordinate wave function of one and the same bound state independently
of its spin symmetry:  $Q_{X}(\q)=Q_{Y}(\q)=A(\q)$;
$K^{(\PP)}_Z([\q_1\q_2\q_3])=0$;
$ K^{(\PP)}_Y([\q_1 \q_2]\q_3)= A(\q_1)-A(\q_2)$;
$K^{(\PP)}_X(\{\q_1\q_2\}\q_3)= A(\q_1)+A(\q_2)- 2A(\q_3)$;
where $A(\q)$ satisfies the eq. (\ref{QDY}).
As was shown in \cite{Shond3}, \cite{MMM}, the asymptotic behavior
(\ref{ATgn}) guarantees that for both cases we deal with self-adjoint
semi-bounded bellow three-particle Hamiltonian.
Whereas the Hamiltonians corresponding to more slowly vanishing T-matrix
of another two-particle extensions (\ref{FB}) are unbounded and correspond
to "collapse" in three-particle system.

If for $\PP=0$ we consider zero orbital momentum subspace,
$A(\q)\Longrightarrow A(q)$, the corresponding equations read
\bea
&& q\, A(q)=T(\varrho (q))\frac{\xi}{\pi}\int^\infty_0 d k\, k\, A(k)
\ln\left(\frac{k^2+q^2+kq+\omega^2}{k^2+q^2-kq+\omega^2}\right),
\label{A0}
\eea
where $\xi=2$ for the case 1) and $\xi=-1$ for the case 2). A simple analysis,
curried out in Appendix,
shows that for appropriate conditions the integral operator written here is
equivalent to symmetrical quite continuous positively defined operator of
Hilbert-Schmidt type.
Therefore nontrivial solutions of (\ref{A0}) may be take place only for
case 1). Whereas for case 2) bound states are impossible.
We conclude that the bound states of three identical particle may be appear
in this model only for isospin 1/2 with $X$-type wave functions (\ref{K1/2X}).

\begin{center}
{\bf 5. Conclusions}
\end{center}

So, in \cite{YdFz} and here we formulate unambiguous renormalization
procedure to extract a renormalized dynamic from "nonrenormalizable"
contact four-fermion interaction that is selfconsistent
in every $N$-particle sector and is intimately connected with construction
of self-adjoint extension of corresponding quantum mechanical Hamiltonian
and restoration of Galileo invariance. We have shown that the simple
$\Lambda$-cut-off prescription with definite $\Lambda$ dependence of "bare"
quantities and fine-tuning relations lead for reduced field Hamiltonian
(\ref{aH0}) to the set of self-adjoint semi-bounded Hamiltonians in one-,
two- and three-particle sectors with correctly defined solutions for
scattering and bound states.

The authors are grateful to A.A.Andrianov, V.B.Belyaev and W.Sandhas for
useful remarks.
\begin{center}
{\bf Appendix}
\end{center}

Using hyperbolic substitutions and natural odd continuation onto
a whole real axis for the function
\bea
&& F(q)\equiv \frac{qA(q)}{T(\varrho (q))}=-F(-q)=\varphi(\vartheta)=
-\varphi(-\vartheta);\;\;\;q=\frac{2}{\sqrt{3}}\ \omega\ \sinh\vartheta,
\label{subst1} \\
&& \varrho(q)=\sqrt{\frac{3}{4}q^2+\omega^2}=\omega\ \cosh\vartheta,\;\;
k=\frac{2}{\sqrt{3}}\ \omega\ \sinh\tau,\;\;\varrho(k)=\omega\ \cosh\tau,
\nonumber
\eea
the equation (\ref{A0}) may be reduced to the following convenient form:
\be
\varphi(\vartheta)=\frac{2\xi K}{\pi\sqrt{3}}\di\limits^{\infty}_{-\infty}
d\tau\ W(\cosh\tau)\varphi(\tau)
\ln\left(\frac{2\cosh(\tau-\vartheta)+1}{2\cosh(\tau-\vartheta)-1}\right);
\;\;W(\cosh\tau)=\frac{\omega}{K}\cosh\tau\  T\left( \omega\ \cosh\tau\right),
\label{phieq}
\ee
where $W(\cosh\tau)$ is even function of $\tau$ and
$K$ is appropriate positive constant introduced for convenience.
Note that the last kernel has additional eigenfunctions with opposite parity.

According to general restrictions from two- and three-particle problem
\cite{Merk} we suppose $\Upsilon > 0$, $\omega>b\geq 0$. Therefore, from
(\ref{ATgn}) $T(\varrho (q))>0$ and tends to zero fast enough to make
meaningful the next substitution:
\bea
\vartheta=\vartheta(\eta),\;\;\;\tau=\tau(\zeta),\;\;\;
d\zeta=d\tau W(\cosh\tau),\;\;\;\varphi(\vartheta)=\Phi(\eta)=-\Phi(-\eta);
\;\mbox{ where}
\nonumber \\
\zeta(\tau)=-\zeta(-\tau)=\di\limits_{-\infty}^{\tau}dr W(\cosh r)-\chi;
\;\;\; 2\chi\equiv\di\limits_{-\infty}^{\infty}dr W(\cosh r); \;\;\;
\infty >2\chi >0.
\label{subst2}
\eea
It is obviously true for arbitrary $T(\varrho (q))$ with the above properties
and it transforms (\ref{phieq}) to equation with symmetrical quite continuous
kernel:
\be
\Phi(\eta)=\frac{4\xi K}{\pi\sqrt{3}} \di\limits^{\chi}_{-\chi}d\zeta\
\Phi(\zeta)\ \ln\left[\frac{2\cosh\left(\tau(\zeta)-\vartheta(\eta)\right)+1}
{2\cosh\left(\tau(\zeta)-\vartheta(\eta)\right)-1}\right]\equiv
\frac{4\xi K}{\pi\sqrt{3}} \left(\hat{\cal L}\Phi\right)(\eta).
\label{symeq}
\ee
With the usual definition of scalar product in $L_2(-\chi,\chi)$ for
arbitrary function $\Phi(\eta)$ from this space one has, using Fourier
transformation:
\be
\left(\hat{\cal L}\Phi,\Phi \right)=\di\limits^{\infty}_{-\infty}d\nu
\left| g(\nu)\right|^2\,\frac{\sinh(\pi\nu/6)}{\nu\cosh(\pi\nu/2)} > 0;
\;\;\;g(\nu)\equiv\di\limits^{\infty}_{-\infty}d\tau\,e^{i\nu\tau}\,
W(\cosh\tau)\,\varphi(\tau).
\ee
Therefore, all eigenvalues of operator $\hat{\cal L}$ are positive.

At last, for $b=0$, $\Upsilon=\omega\ \cosh 2\chi$, $K=2\coth2\chi$,
from (\ref{subst2}) follow the manifest expressions
\be
e^{\vartheta}=\frac{\sinh(\chi+\eta)}{\sinh(\chi-\eta)},\;\;
e^{\tau}=\frac{\sinh(\chi+\zeta)}{\sinh(\chi-\zeta)};\;\;\;
e^{2\eta}=\frac{\cosh(\chi+\vartheta/2)}{\cosh(\chi-\vartheta/2)},\;\;
e^{2\zeta}=\frac{\cosh(\chi+\tau/2)}{\cosh(\chi-\tau/2)};
\label{nxtsub}
\ee
which allow direct application of Faddeev consideration \cite{Merk} to eq.
(\ref{symeq}) when $\omega\rightarrow 0$, $\chi\rightarrow\infty$. Thus
$\cosh (\tau-\vartheta)\simeq \cosh 2(\zeta-\eta)$ and the seeking of
coefficients $a_m$ of Fourier expansion
\[
\Phi(\eta)=\sum^{\infty}_{m=-\infty}a_m\ e^{i\pi m\eta/\chi};\;\;\;
a_m=\sum^{\infty}_{n=-\infty}{\cal C}^{\chi}_{mn}\ a_n;\;\;\;
{\cal C}^{\chi}_{mn}=\frac{4\xi K}{\pi\sqrt{3}}
\left(\hat{\cal L}e^{i\pi n\zeta/\chi}, e^{i\pi m\eta/\chi}\right),
\]
leads to Faddeev's like relation:
\[ 1=\frac{2\xi K}{\pi\sqrt{3}}\,
\frac{\sinh(\pi\nu/6)}{\nu\cosh(\pi\nu/2)};\;\;\;\nu\equiv\frac{\pi m}{2\chi},
\]
which has a solution $\nu=\nu_0>0$ indead only for the case 1)
$(\xi=2)$, giving asymptotic distribution of Efimov levels as:
$\omega_m\simeq\ 2\Upsilon e^{-\pi m/\nu_0}$.


\begin{thebibliography}{99}
\bibitem{ALSHT}A.N.\,Vall, et.al. in D.V. Shirkov, D.I. Kazakov,
A.A.Vladimirov, editors, {\it Proceedings of X International Conference on
Problems of Quantum Field Theory}, JINR E2-96-369, p. 214, Dubna, 1996.
\vspace{-2.5mm}
\bibitem{YdFz}A.N.\,Vall, et.al., YaF (Russ.Jour. of Nucl.Phys.), 1997,
{\bf 60}, N7.
\vspace{-2.5mm}
\bibitem{dnmpZ} A.N.\,Vall, S.E.\,Korenblit, V.M.\,Leviant, A.B.\,Tanaev,
in B.B.\,Levchenko, V.I.\,Savrin, editors,
{\it Proceedings of X-th International WorkShop on High Energy Physics and
Quantum Field Theory}, p. 279, Moscow State University, 1996.
\vspace{-2.5mm}
\bibitem{BrFdd}F.A.\,Berezin, L.D.\,Faddeev, Dokl. Akad. Nauk SSSR, 1961,
{\bf 137}, 1011.
\vspace{-2.5mm}
\bibitem{YShr}Yu.M.\,Shirokov, TMF (Sov. Jour. of Theor. Math. Phys.),
1980, {\bf 42}, 45; 1981, {\bf 46}, 291; 1981, {\bf 46}, 310;
\vspace{-2.5mm}
\bibitem{Shond}Yu.G.\,Shondin, TMF (Sov. Jour. of Theor. Math. Phys.),
1985, {\bf 64}, 432; 1985, {\bf 65}, 24; 1988, {\bf 74}, 331.
\vspace{-2.5mm}
\bibitem{Fewst}C.J.\,Fewster, J.Phys.A: Math. Gen. 1995 {\bf 28}, 1107.
J.F. van Diejen, A.Tip, J.Math.Phys. 1991, {\bf 32}, (3), 630.
\vspace{-2.5mm}
\bibitem{Ber}F.A.\,Berezin, Math.Collection, 1963, {\bf 60}, 425.
\vspace{-2.5mm}
\bibitem{Thorn}C.\,Thorn, Phys. Rev. 1979, {\bf D19}, 639.
\vspace{-2.5mm}
\bibitem{Shond3}Yu.G.\,Shondin, TMF (Sov. Jour. of Theor. Math. Phys.),
1982, {\bf 51}, 181.
\vspace{-2.5mm}
\bibitem{MMM}K.A.\, Makarov, V.V.\, Melejik, A.K.\,Motovilov, Preprint, JINR,
P5-94-51, Dubna, 1994; K.A.\, Makarov, V.V.\, Melejik,
TMF (Sov. Jour. of Theor. Math. Phys.), 1996, {\bf 107}, 415.
\vspace{-2.5mm}
\bibitem{Merk}S.P.\,Merkuryev, L.D.Faddeev, Quantum Scattering Theory for
few-body systems, "Nauka", 1985.
\end{thebibliography}
\end{document}